 \documentclass[aps, twocolumn, showpacs, amsmath]{revtex4-1}
\usepackage{dcolumn}
\usepackage{bm}
\usepackage{graphicx}
\usepackage{color}
\usepackage{subfigure}
\usepackage{hyperref}
\usepackage{latexsym}
\usepackage{amsthm}
\usepackage{amssymb}
\usepackage{braket}
\DeclareGraphicsExtensions{.jpg,.pdf, .mps, .png, .eps, .ps, .EPS,.gif}

\begin{document}
\def\be{\begin{equation}}
\def\ee{\end{equation}}

\def\bc{\begin{center}}
\def\ec{\end{center}}
\def\bea{\begin{eqnarray}}
\def\eea{\end{eqnarray}}
\newcommand{\avg}[1]{\langle{#1}\rangle}
\newcommand{\Avg}[1]{\left\langle{#1}\right\rangle}

\def\ie{\textit{i.e.}}
\def\etal{\textit{et al.}}
\def\m{\vec{m}}
\def\G{\mathcal{G}}

\newcommand{\davide}[1]{{\bf\color{blue}#1}}
\newcommand{\gin}[1]{{\bf\color{green}#1}}

\title{Percolation on  branching   simplicial and cell complexes  \\and its relation to  interdependent percolation}

\author{Ginestra Bianconi}
\affiliation{
School of Mathematical Sciences, Queen Mary University of London, London, E1 4NS, United Kingdom\\The Alan Turing Institute, The British Library, London NW1 2DB, United Kingdom
}
\author{Ivan Kryven}
\affiliation{Mathematical Institute, Utrecht University, PO Box 80010, 3508 TA Utrecht,  the Netherlands}
\author{Robert M. Ziff}
\affiliation{Center for the Study of Complex Systems and Department of Chemical Engineering, University of Michigan, Ann Arbor, Michigan 48109-2800, USA}

\begin{abstract}
Network geometry has strong effects on network dynamics. In particular, the underlying hyperbolic geometry of discrete manifolds has  recently been shown to affect their critical percolation properties.
Here we  investigate the properties of link percolation in non-amenable two-dimensional branching  simplicial and cell complexes, i.e., simplicial and cell complexes in which the boundary scales like the volume.  We establish the relation between the equations determining the percolation probability in random branching cell complexes and the equation for interdependent percolation in multiplex networks with inter-layer degree correlation equal to one. By using this relation we show that branching cell complexes can display more than two percolation phase transitions:  the  upper percolation transition, the lower  percolation transition, and one or more intermediate phase transitions.  At these additional transitions the percolation probability and the fractal exponent both feature a discontinuity. Furthermore, by using the renormalization group theory we  show that the upper percolation transition can belong to various universality classes including the Berezinskii-Kosterlitz-Thouless (BKT) transition, the discontinuous percolation transition, and continuous transitions with anomalous singular behavior that generalize the BKT transition.
\end{abstract}


\maketitle

\section{Introduction}

Understanding the interplay between network structure  and dynamics \cite{Doro_crit} has been a fundamental research question in statistical mechanics of networks \cite{Netsci,dorogovtsev2013evolution}. Recently this field has gained much momentum thanks to the vibrant research on generalized network structures, including multilayer networks \cite{bianconi2018multilayer,PhysReport,Kivela} and simplicial and cell complexes \cite{Perspectives,Lambiotte,Emergent,NGF,Polytopes,Bianconi_Ziff,kryven2019renormalization,millan2018complex,millan2019synchronization,vsuvakov2018hidden,
iacopini2019simplicial,skardal2019abrupt}.  In particular the study of percolation \cite{Ziff_Review,Kahng_review} on generalized network structures has renewed interest to the so-called {\em explosive phenomena} \cite{Explosive_phenomena}, which have been investigated for single networks \cite{dSouza,Riordan,Doro_explosive,Ziff_explosive,Ldev} as well as in the context of interdependent percolation in multilayer networks \cite{Havlin,bianconi2018multilayer,Havlin2,Doro_multiplex,kryven2019bond,bianconi2014} and  percolation in hyperbolic simplicial and cell complexes \cite{hyperbolic_Ziff,Bianconi_Ziff,kryven2019renormalization}.
Multilayer networks include links representing interactions of different natures and connotations. As such, multilayer networks can describe interacting networks  as diverse as global infrastructures, financial systems, and  the brain.
 In the last years it has been shown \cite{Havlin,bianconi2018multilayer} that interdependent percolation of multilayer networks leads to discontinuous phase transitions, revealing their intrinsic fragility.
Simplicial and cell complexes are built using geometrical building blocks comprising of triangles, polygons and polytopes. As such they are  ideal generalized network structures to investigate the interplay between hyperbolic network geometry and dynamics
 \cite{hyperbolic_Ziff,Bianconi_Ziff,kryven2019renormalization,Gu_Ziff,Moore_Mertens,Emergent,percolation_Apollonian}. Recently it has been found \cite{hyperbolic_Ziff} that link percolation in hyperbolic Farey graphs and  some well-behaved two-dimensional hyperbolic manifolds constituting the skeleton of cell complexes \cite{kryven2019renormalization} is discontinuous. 
Despite the fact that both  percolation on simplicial and cell complexes and interdependent percolation in multilayer networks can lead to discontinuous phase transitions, the relation between the two critical phenomena has not been so far investigated. 

In this paper  we  depart from the study of discrete  manifolds and we consider branching simplicial and cell complexes that reduce in some limit to very well-studied hierarchical network structures, as for instance, the flower network and its generalizations \cite{Patchy,flower_tau,tricritical,clusters}. These branching simplicial and cell complexes  display a  critical behavior of percolation  that can be  fully characterized  using the renormalization group (RG)   \cite{RG,Boettcher_RG, Boettcher_Potts, Berker_RG,kryven2019renormalization,doro2,doro3}. 
Interestingly, our RG investigation  of branching simplicial and cell complexes reveals a surprising relation between percolation in these structures  and interdependent percolation on multilayer networks.
Namely,  we uncover  a mathematical  mapping between the equation determining the percolation probability in some specific  simplicial and cell complexes and the equation determining the emergence of the Mutually Connected Giant Component (MCGC) \cite{Havlin,bianconi2018multilayer} of  correlated and interdependent multiplex networks \cite{kryven2019sharp}.
These correlated multiplex networks have  recently been shown to be able to sustain multiple phase transitions \cite{kryven2019sharp}.
Building upon the revealed  mathematical mapping between the percolation in branching cell complexes and interdependent percolation, we are able to show that branching simplicial and cell complexes are not only characterized by their upper and lower percolation thresholds, as is the general rule for all non-amenable graphs \cite{Lyons}, but they can feature intermediate phase transitions as well. These intermediate percolation transitions are discontinuous and can be observed when the distribution $r_k$, which denotes the probability that a randomly chosen link branches into $k$ $m$-polygons, is multi-modal.

Moreover, in this paper we  identify  the conditions that guarantee a non-trivial discontinuous percolation transition at the upper percolation threshold of two-dimensional simplicial and cell complexes. Using the RG technique we show that as the topology of the branching simplicial and cell complexes changes, it is possible to observe a change of universality class of percolation between the discontinuous and Berezinskii-Kosterliz-Thouless (BKT) phase transitions, confirming and generalizing the results in Refs.\ \cite{Patchy,flower_tau,tricritical,clusters}. Moreover, we show that  the system might display higher-order critical points corresponding  to continuous transitions with non-trivial singular behavior, which, to our knowledge, has not been reported previously on similar structures.

The paper is structured as follows:
in Sec.\ II we define the branching simplicial and cell complexes considered in this paper; in Sec.\ III we characterize their percolation probability; in Sec.\ IV we reveal the relation with interdependent percolation of correlated multiplex networks; in Sec.\ V we show that simplicial and cell complexes can undergo more than two percolation phase transitions; in Sec.\ VI we derive the expression of the generating function of the cluster-size distribution; in Sec.\ VII we derive the expression and the critical behavior of the fractal exponent, and in Sec.\ VIII we use the RG approach to predict the nature of the percolation transition at the upper percolation threshold. Finally in Sec.\ IX we provide the conclusions.
 \begin{figure}
\begin{center}
\includegraphics[width=\columnwidth]{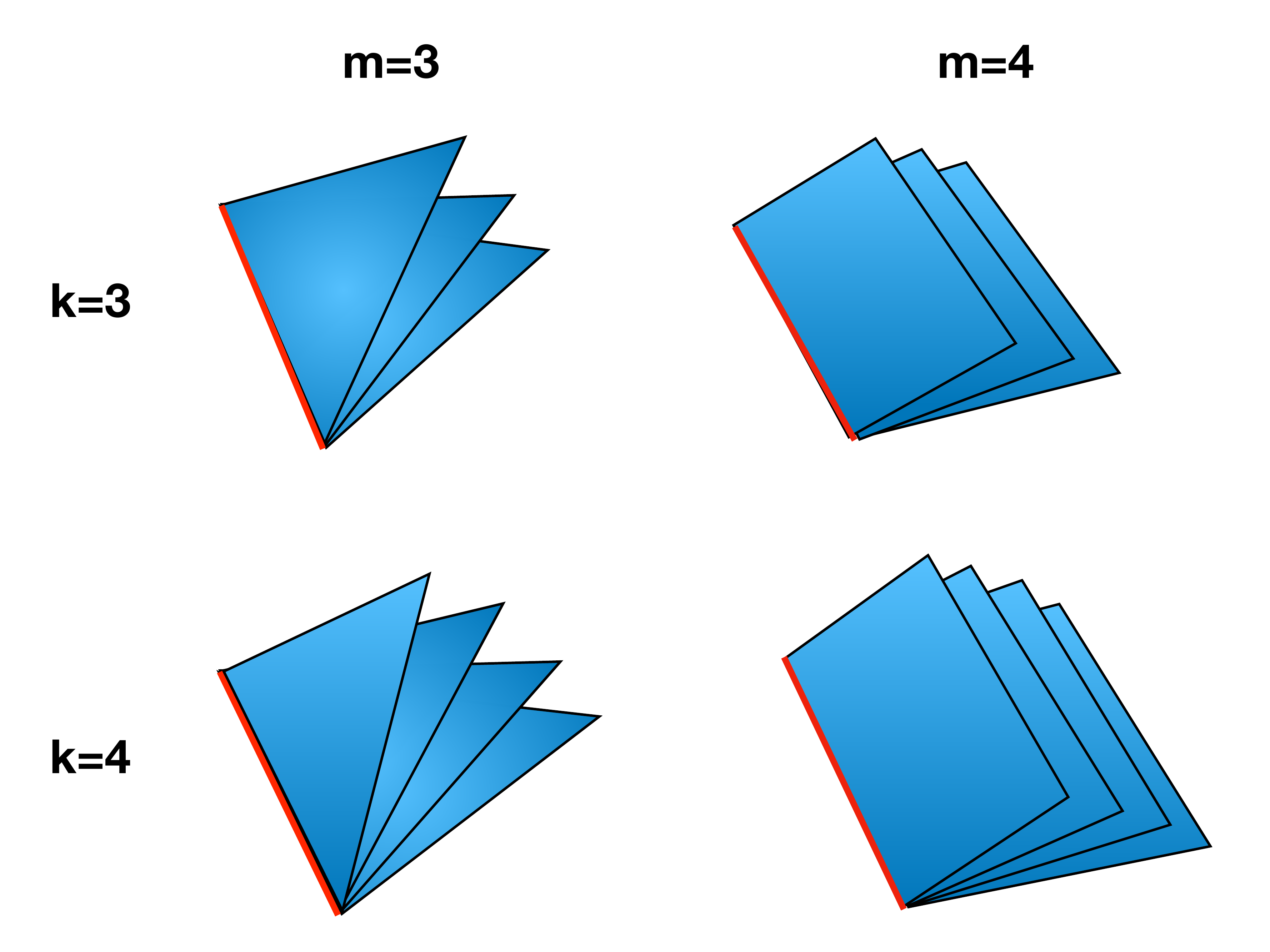}
\caption{We show the first iteration $n=1$  of the branching process in which the initial single link (shown as a thick red line) branches out to $k$, $m$-polygons with $k$ drawn from the distribution $r_k$. Here the first iteration is shown  for different values of $m=3,4$ (corresponding to the attachment of triangles and rectangles respectively) and $k=3,4$.}
\label{fig.book}
\end{center}
\end{figure}
  \begin{figure}
 \begin{center}
\includegraphics[width=\columnwidth]{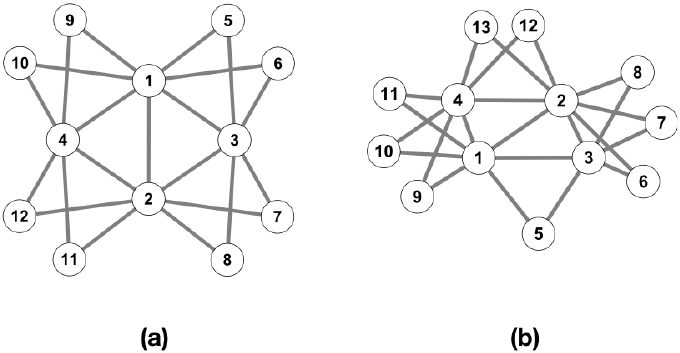}
\caption{The result of the first $n=2$ iterations are shown for: (a) the flower network with $m=3$ and $r_k=\delta_{k,2}$, and (b) for a random branching cell complex with $m=3$ with non-zero probabilities $r_1,r_2,r_3$.  In both cases nodes $1,2$ are the nodes present at the iteration $n=0$, nodes $3,4$ are the nodes added at the iteration $n=1$, and all the other nodes are added at iteration $n=2$. }
\label{fig.flower}
\end{center}
\end{figure}

\section{Branching simplicial and cell complexes}

Hierarchical networks with a non-amenable structure can be manifolds, or more generally, branching simplicial and cell complexes. Here we focus on a few specific examples of two-dimensional branching simplicial and cell complexes.
A two-dimensional simplicial complex is a topological structure formed by gluing triangles along their links, whereas a two-dimensional cell complex generalizes this concept by gluing arbitrary $m$-polygons along their links.
Two-dimensional simplicial and cell complexes form manifolds only if each link is incident to at most two triangles (for the simplicial complexes) or two polygons (for the cell complexes).
However, if this condition is not satisfied, the resulting topological structures  are not discrete manifolds.
A branching hierarchical simplicial complex or cell complex is a network that can be constructed by successively gluing triangles or polygons to single links such that each link can be incident to more than two polygons.
Probably the most discussed hierarchical branching simplicial complex is the flower network which stems from the Migdal-Kadanoff \cite{migdal1976phase,kadanoff1976notes}
  renormalization group techniques.
Here we focus on the percolation properties  of   branching simplicial and cell complexes that feature random and deterministic structure and generalize the flower network as described below.

 A branching tree can be constructed by starting at iteration $n=0$ from a single node. At every time $n>0$ the node connected to a single link is connected to $k$ new links with probability $r_k$.
 This construction can be generalized to random branching cell complexes in dimension $d=2$.
 We start at iteration $n=0$ from a single initial link.
 At iteration $n=1$ we attach  a $k\geq 1$ polygons with $m$ faces ($m$-polygons) to it with $k$ drawn from the probability distribution $r_k$ (see Fig.~$\ref{fig.book}$).
At iteration $n>1$,  we glue  $k\geq 1$ new $m$-polygons with probability $r_k$ to each link introduced at iteration $n-1$.
 In this way, the number of polygons $k$ is treated as a random variable. 
 
At iteration $n$ the average number of nodes $\bar{N}_n$ and links $\bar{L}_n$  is given by 
\bea\label{barN}
\bar{N}_n&=&2+\frac{\avg{k}(m-2)}{\avg{k}(m-1)-1}[\avg{k}^n(m-1)^n-1],\\
\bar{L}_n&=&\frac{1}{\avg{k}(m-1)-1}[\avg{k}^{n+1}(m-1)^{n+1}-1].
\eea
In the case  $m=3$  the polygons are triangles, and the same iterative process generates a random branching simplicial complex.
Here we refer to networks with arbitrary fixed $m\geq 3$ as simply the cell complexes.
 We note that the case $m=3$ and $r_k=\delta_{k,2}$ corresponds to the flower network shown in Fig.~$\ref{fig.flower}$a. A branching cell complex with $m=3$ and heterogeneous $r_k$ distribution is shown in Fig.~$\ref{fig.flower}$b.


\section{Percolation in non-amenable network structures}
We consider link percolation on branching cell complexes where we remove  each link  independently  with probability $q=1-p$.
Since the random branching cell complexes that we consider are non-amenable structures \cite{Lyons}, link percolation displays   at least two percolation thresholds.
In particular, as in hyperbolic manifolds   \cite{hyperbolic_Ziff,Bianconi_Ziff,kryven2019renormalization}, we distinguish between  the lower $p^{\star}$ and the upper $p_c$ percolation thresholds leading to the identification of    three distinct phases.
\begin{itemize}
\item[(1)] For $p<p^{\star}$, i.e., below the lower percolation threshold, there is no infinite cluster. Therefore the percolation probability, i.e., the probability that the two initial nodes are connected at least by a path of non-damaged nodes, is $T=0$. 

\item[(3)] For $p^{\star}<p<p_c$ the percolation probability $0<T<1$ and $M_n$, the number of nodes in the largest component at iteration $n$,  is subextensive, i.e., 
\bea
M_n=\left(\bar{N}_n\right)^{\psi_n},
\label{Mn}
\eea
where the limit of the exponent $\psi_n$ for $n\to \infty$, i.e.,
\bea
\psi=\lim_{n\to \infty}\psi_n
\label{psi_def}
\eea
is called the {\em fractal exponent}. In this phase we have $\psi<1$.
\item[(3)] For $p>p_c$, i.e., above the upper percolation threshold,   there is an infinite cluster which is extensive. This implies that the percolation probability $T=1$ and $\psi=1$. Moreover, if we indicate with $M_n$ the number of nodes in the largest component at iteration $n$ above the upper percolation threshold, the fraction of nodes in the largest component $P_{\infty}(p)$ is of order one, i.e.,
\bea
P_{\infty}(p)=\lim_{n\to \infty}\frac{M_n}{\bar{N}_n}={\it O}(1).
\label{Pinfty0}
\eea
\end{itemize}

\section{Percolation probability}

In this section we investigate the percolation probability $T_n$ indicating the probability that the two nodes present at iteration $n=0$ are connected with a path within the first $n$ iterations.

The  percolation probability $T_n$ for the random branching cell complexes satisfies the following recursive equation
\bea
T_{n+1}=1-(1-p)\sum_{k\geq 1} r_k\left(1-T_n^{m-1}\right)^k,
\label{RG}
\eea
That is, the two initial nodes are not connected at iteration $n+1$ if the link that connects the initial nodes is removed and there is no path connecting these nodes through the $m$-polygons within the first $n$ iterations.
In the limit of an infinite network, $n\to \infty$, the linking probability $T_n$ converges to the percolation probability $T$, i.e., $T_n\to T$, where $T$ satisfies 
\bea
T=1-(1-p)\sum_{k\geq 1} r_k\left(1-T^{m-1}\right)^k.
\label{T}
\eea
Note that  in presence of multiple solutions of Eq.\ (\ref{T}) we only consider the solution $T\in [0,1]$ with the smallest value.  
By defining $R(z)$ as the generating function of the distribution $r_k$, i.e.,
\bea\label{R}
R(z)=\sum_{k\geq 1} r_k z^k,
\eea
and by defining 
$Q(T)$ as
\bea
\label{QT}
Q(T)=T^{m-1},
\eea 
we can write Eq.\ (\ref{T}) as 
\bea
f(T)=T-1+(1-p)R(1-Q(T))=0.
\label{FT}
\eea
A close inspection of this equation reveals important properties of the critical behavior of the random branching cell complexes.
We first  observe that $T=0$ is a solution of Eq.\ (\ref{FT}) only for $p=0$. Therefore it follows that the lower percolation threshold $p^{\star}$ is given by 
\bea
p^{\star}=0,
\eea
for every branching cell complex studied in this work.

Secondly we will show that  applying the theory of critical phenomena, we can obtain information on the other possible percolation thresholds that can be encountered at  discontinuous hybrid critical points, second-order critical points, and higher-order critical points of Eq.\ (\ref{FT}).
The discontinuous hybrid critical point, also called the {\em saddle-node bifurcation}, is found  
by imposing $T_c<1$ and
\bea
\left.f(T_c)\right|_{p=p_c}&=&0,\nonumber\\
\left.f^{\prime}(T_c)\right|_{p=p_c}&=&0,\nonumber\\
\left.f^{\prime\prime}(T_c)\right|_{p=p_c}&<&0.
\label{cond_disc}
\eea
If the above equations yield many solutions $(p_c,T_c)$ one has to select a subset that forms a minimal subsequence that is simultaneously ascending in both variables $p_c$ and $T_c$.
In order to find the critical behavior of $\Delta T$ let us  expand Eq.\ (\ref{FT}) close to this critical point  characterized by $|\Delta T|=|T-T_c|\ll1 $ and $|\Delta p|=|p-p_c|\ll1$ and $\Delta p<0$, $\Delta T<0$.
In this way we get
\bea
0&=&\left. f(T_c)\right|_{p=p_c}+\left. f^{\prime}(T_c)\right|_{p=p_c}\Delta T+\left.\frac{\partial f(T_c)}{\partial p}\right|_{p=p_c}\Delta p \nonumber \\
&&+\left.\frac{\partial^2 f(T_c)}{\partial p\partial T}\right|_{p=p_c}
\Delta p\Delta T+\frac{1}{2}\left.\frac{\partial^2 f(T_c)}{\partial T^2}\right|_{p=p_c}(\Delta T)^2+\ldots\nonumber
\eea

Since $T_c<1$ we have $\left.{\partial f(T_c)}/{\partial p}\right|_{p=p_c}<0$. Therefore, for  $|\Delta T|=|T-T_c|\ll1 $ and $|\Delta p|=|p-p_c|\ll1$ with  $\Delta p<0,\Delta T<0$, we have the hybrid critical behavior 
\bea
\hat{A} \Delta p \simeq  (\Delta T)^{2},
\eea
or equivalently
\bea
|\Delta T| \simeq |\hat{A} \Delta p|^{{\beta}},
\eea
where \bea
{\beta}=\frac{1}{2},
\eea
and 
\bea
\hat{A}&=&-\left[\left.\frac{\partial f(T_c)}{\partial p}\right|_{p=p_c}\right]\left[\frac{1}{2}\left.\frac{\partial^2 f(T_c)}{\partial T^2}\right|_{p=p_c}\right]^{-1}.
\eea
The second-order critical point, also called the {\em transcritical bifurcation}, is characterized by  the following conditions on $f(T)$
\bea
\left.f(1)\right|_{p=p_c}&=&0,\nonumber \\
\left.f^{\prime}(1)\right|_{p=p_c}&=&0,\nonumber \\
\left.f^{\prime\prime}(1)\right|_{p=p_c}&<&0.
\label{cond}
\eea
In this way we find the critical point $(p_c,T_c)$ with 
\bea
p_c&=&1-\frac{1}{(m-1)r_1},\nonumber \\
T_c&=&1,
\label{2pc}
\eea
obtained as long as there is no discontinuous critical point for $p<1-1/[(m-1)r_1]$ and provided that
\bea
r_1>\max\left[\frac{1}{m-1},2\frac{m-1}{m-2}r_2\right].
\label{r1}
\eea
In order to find the critical behavior of $\Delta T$ let us  expand Eq.\ (\ref{FT}) close to this critical point  characterized by $|\Delta T|=|T-1|\ll1 $ and $|\Delta p|=|p-p_c|\ll1$ and $\Delta p<0$, $\Delta T<0$.
 By taking into account the critical point conditions expressed in Eq.\ (\ref{cond})  and the fact that  $\left.{\partial f(T_c)}/{\partial p}\right|_{p=p_c}=0$ because $T_c=1$, for $\Delta p<0$ the expansion gives the  mean-field behavior 
\bea
\Delta T \simeq (A \Delta p)^{\beta},
\eea
where
\bea
\beta=1,
\eea
and
\bea
{A}&=&
-\left[\left.\frac{\partial^2 f(T_c)}{\partial p\partial T}\right|_{p=p_c}\right]\left[\frac{1}{2}\left.\frac{\partial^2 f(T_c)}{\partial T^2}\right|_{p=p_c}\right]^{-1}.
\eea

In a model of random branching cell complexes with arbitrary distribution $r_k$, the manifold of second-order critical points meets the manifolds of hybrid transitions on a set of  tricritical points. The tricritical points of Eq.\ (\ref{FT}), also called the {\em pitchfork bifurcation} points,  can be found by imposing the conditions
\bea
\left.f(1)\right|_{p=p_c}&=&0,\nonumber \\
\left.f^{\prime}(1)\right|_{p=p_c}&=&0,\nonumber\\
\left.f^{\prime\prime}(1)\right|_{p=p_c}&=&0,\nonumber \\
\left.f^{\prime\prime\prime}(1)\right|_{p=p_c}&>&0.
\eea
These equation identify 
the tricritical point $(p_c,T_c)$, which is given by 
\bea
p_c&=&1-\frac{1}{(m-1)r_1},\nonumber \\
T_c&=&1,
\eea
 as long as there is no discontinuous transition for $p<1-1/[{(m-1)r_1}]$ and provided that
\bea
r_1&=&2\frac{m-1}{m-2} r_2
\eea
where
\bea
r_1>\max\left[\frac{1}{m-1},6\frac{(m-1)^2}{(2m-3)(m-2)}r_3\right].
\eea

We expand Eq.\ (\ref{FT})  close to the critical point for $|\Delta T|=|T-1|\ll1 $ and $|\Delta p|=|p-p_c|\ll1$ up to third order. We observe that since $T_c=1$ we have $\left.{\partial f(T_c)}/{\partial p}\right|_{p=p_c}=0$. Therefore  for $\Delta p<0$ we obtain the tricritical scaling
\bea
\tilde{A}\Delta p\simeq(\Delta T)^2,
\eea
or equivalently
\bea
|\Delta T| \simeq |\tilde{A} \Delta p|^{{\beta}},
\eea
where \bea
{\beta}=\frac{1}{2},
\eea
and \bea
\tilde{A}&=&
-\left[\left.\frac{\partial^2 f(T_c)}{\partial p\partial T}\right|_{p=p_c}\right]\left[\frac{1}{3!}\left.\frac{\partial^3 f(T_c)}{\partial T^3}\right|_{p=p_c}\right]^{-1}.
\eea
Similarly it is possible to observe even higher-order critical points of order $s>3$.   Such {\em higher-order pitchfork bifurcations} are characterized by  
\bea
\left.f^{(j)}(1)\right|_{p=p_c}&=&0,\nonumber \\
\eea
for $j=0,1,2\ldots, s-1$ and 
\bea
(-1)^{s-1}\left.f^{(s)}(1)\right|_{p=p_c}&>&0.
\eea
Consequently these higher-order critical points are observed when 
\bea
r_k =\frac{\Gamma \left(k-\frac{1}{m-1}\right)}{\Gamma \left(1-\frac{1}{m-1}\right) \Gamma (k+1)} r_1\eea
 for $k=2,\dots,s-1,$ and
 \bea r_{1} > \max\left[\frac{1}{m-1},\frac{\Gamma \left(1-\frac{1}{m-1}\right) \Gamma (s)}{\Gamma \left(s-\frac{1}{m-1}\right)}r_s\right].
 \eea 
 These transitions occur at  $$p_c=1-\frac{1}{(m-1)r_1}$$
 as long as there are no discontinuous critical points for $p<1-1/[{(m-1)r_1}]$.
 At a critical point of order $s$ we observe the critical scaling 
\bea
|\Delta T| \simeq |\tilde{A}_s \Delta p|^{{\beta}},
\eea
where
$$\beta=\frac{1}{s-1},$$ and $\tilde{A}_s$ is given by 
$$
\tilde{A}_s=-\left[\left.\frac{\partial^{2} f(T_c)}{\partial p\partial T}\right|_{p=p_c}\right]\left[\frac{1}{(s+1)!}\left.\frac{\partial^s f(T_c)}{\partial T^s}\right|_{p=p_c}\right]^{-1},
$$
Finally, in the limiting case of $s\to\infty$, the critical exponent $\beta$ vanishes and $T$ features a  $0$-to-$1$ discontinuity at $p_c=0$.
\begin{table}
\begin{tabular}{c c}
Branching Cell-Complex  \qquad & Correlated Multiplex Network\\
 $T$					& $1-S'$\\
$p$					&$1-\tilde{p}$\\
 $m$				& $\kappa+1$\\
 $r_k$				&$P(B)$
\end{tabular}
\caption{Mathematical mapping between the quantities determining the percolation probability $T$ in the modified  branching cell complex and the mathematical quantities determining the probability $S'$ that by following a random link of a random correlated multiplex  network with activity  distribution $P(B)$ and  homogeneous degree of each replica node $\kappa=m-1$ we reach a node in the MCGC.}
\label{table_mapping}
\end{table}

\section{Mathematical mapping to  interdependent percolation in degree-correlated multiplex networks }
The equation determining the percolation probability in random branching cell complexes can be related to the equations determining the MCGC in a correlated multiplex network.
However, in order to perform an exact mathematical mapping between  the two problems and their corresponding equations, one should consider a slight modification of the original model of random branching cell complexes. Consider the modified branching cell complex model with $m\geq 3$ in which we break the symmetry between the $k$ polygons attached to any given link and impose a maximum value  $k_{max}$. The modified model is defined iteratively as in the following.
 We start at iteration $n=0$ from a single initial link.
  At iteration $n\geq 1$ to each link introduced at iteration $n-1$ we glue   $(k-1) $ $m$-polygons and one $(m-1)$-polygon to it with $k\geq 1$ drawn from the  probability distribution $r_k$.
 For this modified random branching cell complex model, the probability that the two initial nodes are connected at iteration $n$ when links are removed with probability $q=1-p$ is given by
 \bea
T_{n+1}=1-(1-p)\sum_{k\geq 1} r_k\left(1-T_n^{m-2}\right)\left(1-T_n^{m-1}\right)^{k-1},\nonumber
\eea
and for large network sizes, when  $n\to \infty$, we have
\bea
T=1-(1-p)\sum_{k\geq 1} r_k \left(1-T^{m-2}\right)\left(1-T^{m-1}\right)^{k-1}.\nonumber
\eea
This equation can be exactly mapped to the equation determining the probability $S^{\prime}$ that by following a link we reach a node in the  MCGC of a multiplex network with heterogeneous activities of the nodes and maximum correlation between the degree of the nodes in different layers.
In fact, let us consider  a multiplex network \cite{bianconi2018multilayer} of $\hat{M}$ layers in which all replica nodes   $(i,\alpha)$ 
in an arbitrary layer $\alpha=1,2\ldots \hat{M}$ have the same activity $B_i=B_{i}^{[\alpha]}$ indicating the number of replica nodes that are interdependent to it and have the same degree $\kappa=\kappa_{i}^{[\alpha]}$ which is independent of the node $i$ and of the layer $\alpha$. In this highly correlated multiplex network studied in Ref.\ \cite{kryven2019sharp}   let us consider the MCGC  when nodes are damaged with probability $1-\tilde{p}$.
The equation for the probability $S'$ that by following a link of the multiplex network in a given layer $\alpha$ we reach a node in the MCGC reads
\bea
S'=\tilde{p}\sum_{B}P(B)[1-(1-S')^{\kappa-1}][1-(1-S')^{\kappa}]^{B-1},\nonumber
\eea
where $P(B)$ is the probability that a random node has activity $B_i=B$. 
At the mathematical level it is thus possible to define a mapping between  the equation determining $T$ in the modified random branching hyperbolic manifolds and the equation determining $S'$ in the correlated multiplex network. 
In this mapping  $m$ corresponds to  $\kappa+1$, the probability distribution $r_k$ corresponds to the activity distribution   $P(B)$ and $p$ corresponds to $1-\tilde{p}$ (see Table \ref{table_mapping}).

Since  the considered multiplex networks  have been shown to display   multiple percolation phase transitions, it follows that the mapping described above suggests that also in the modified random branching network one may expect multiple critical points in the equation determining the linking probability.
However, the question whether also in the originally considered random branching model we can expect multiple critical points of the linking probability needs to be explored in detail and will be addressed in the following section.

\begin{figure}
\begin{center}
\includegraphics[width=0.9\columnwidth]{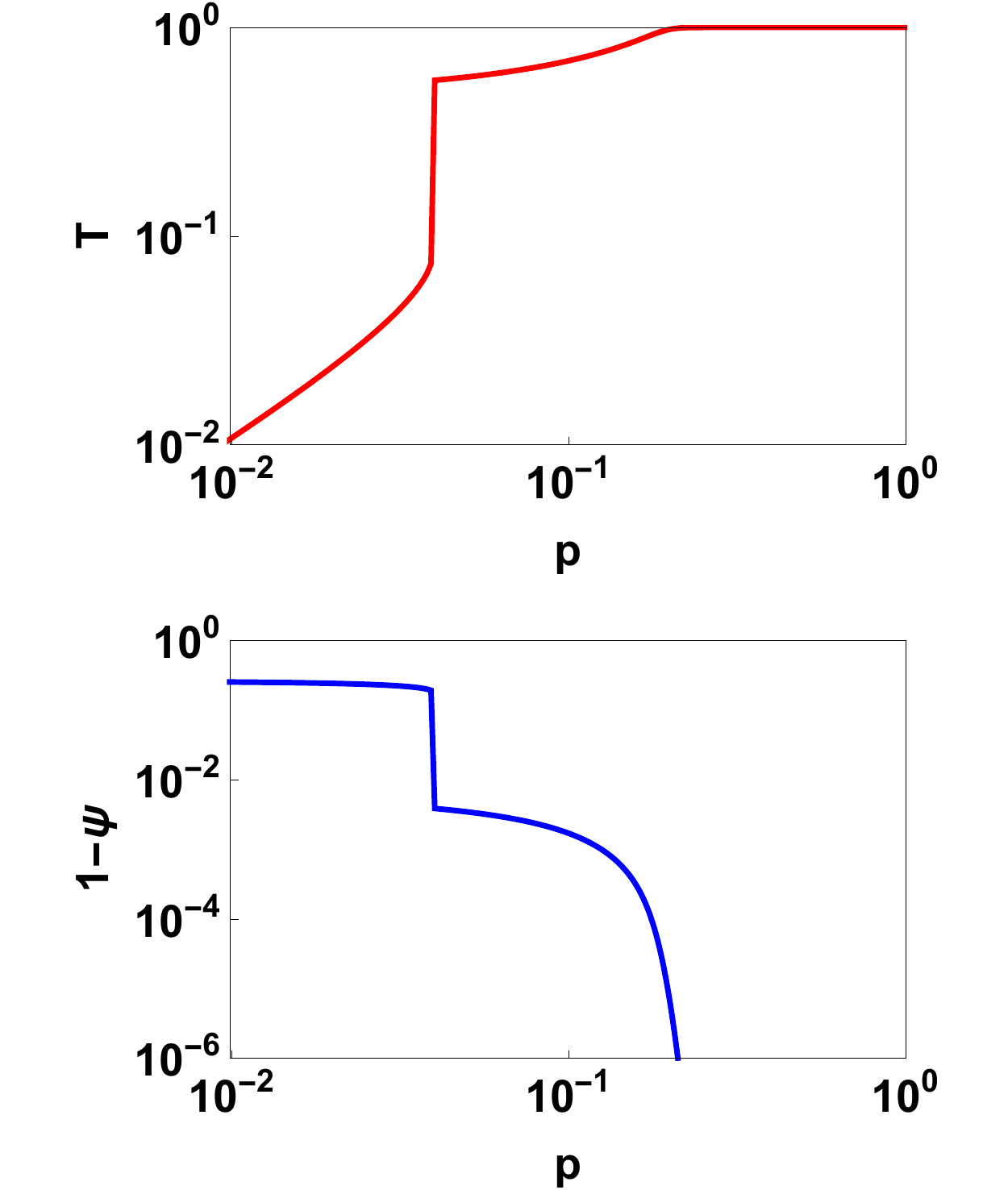}
\caption{The percolation probability $T$ and $1-\psi$ versus the occupation probability of the links $p$,  where $\psi$ is the fractal exponent. The considered  branching simplicial complex ($m=3$) has three percolation thresholds: the lower percolation threshold at $p^{\star}=0$, the upper percolation threshold at $p_c=0.1935(5)$ and one intermediate percolation threshold at $p_c^{\star}=0.0395(6)$. At the intermediate percolation threshold both $T$ and $\psi$ have a discontinuity but remain smaller than one. Here the branching simplicial complex has $r_k=0.62\delta_{k,1}+0.07\delta_{k,2}+0.31\delta_{k,20}.$}
\label{TPsimultiple}
\end{center}
\end{figure}

\begin{figure*}
\begin{center}
\includegraphics[width=\textwidth]{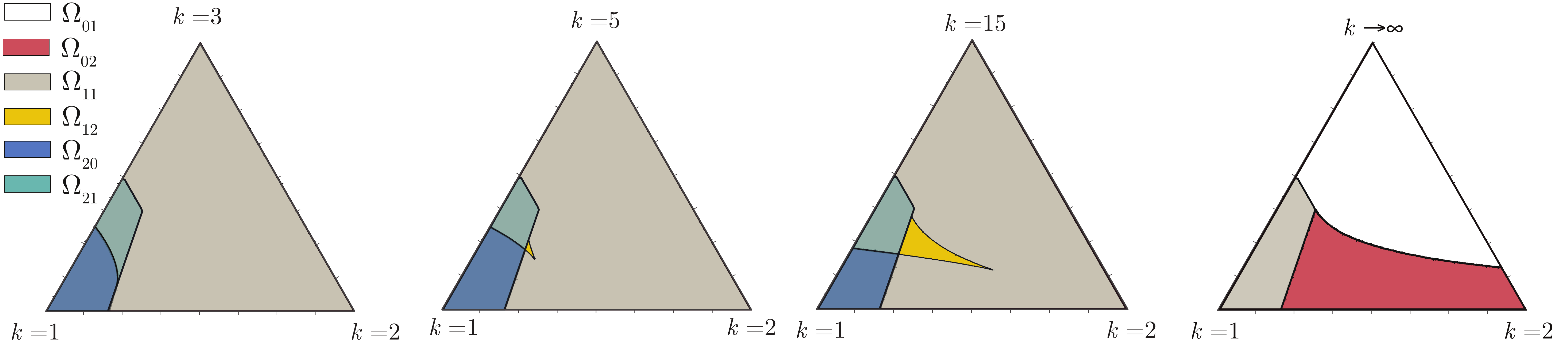}
\caption{The barycentric plot characterizing the phase diagram of link percolation for the branching simplicial complex with $m=3$ and $r_k=r_1\delta_{k,1}+r_{k,2}\delta_{k,2}+r_{\hat{k}}\delta_{k,\hat{k}}$ where $\hat{k}= 3,5,15,$ or is diverging. 
The parameter space $(r_1,r_2,r_{\hat{k}})$ is partitioned into phases  $\Omega_{ij}$ at  which the percolation probability $T$ has  $i$ continuous and $j$ discontinuous transitions.  When $k\to \infty$ in the rightmost barycentric plot, the first two critical points merge and the domains $\Omega_{21},\Omega_{11},\Omega_{12}$ switch correspondingly to $\Omega_{11},\Omega_{01},\Omega_{02}$.
}
\label{fig.combi2}
\end{center}
\end{figure*}

\section{Multiple percolation transitions}

\begin{figure*}
\begin{center}
\includegraphics[width=\textwidth]{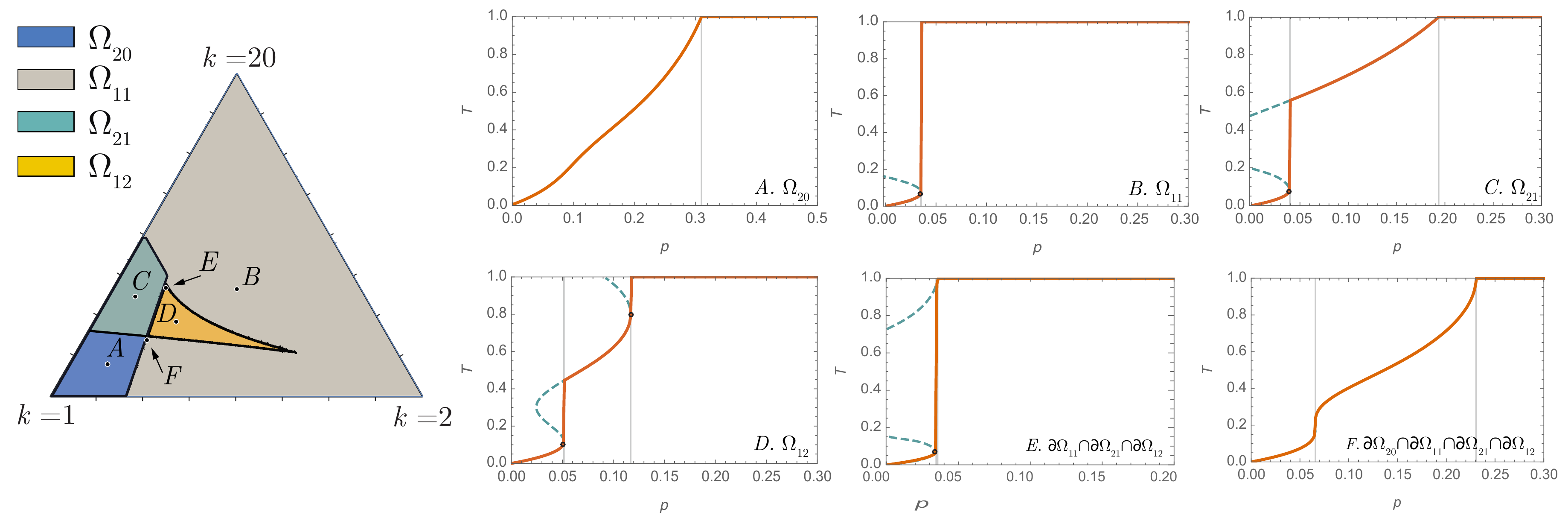}
\caption{ The barycentric plot characterizing the phase diagram of link percolation for the  branching  simplicial complex  with $m=3$ and $r_k=r_1\delta_{k,1}+r_2\delta_{k,2}+r_{20}\delta_{k,20}$.  The percolation probability $T$ versus $p$ is shown at points A,B,C, and D that belong to  phases $\Omega_{20},\Omega_{21},\Omega_{21} $ and $\Omega_{12}$ respectively, at the shared accumulation points $E\in\Omega_{11}\cap\Omega_{21}\cap\Omega_{12}$, and $ F\in\Omega_{20}\cap\Omega_{11}\cap\Omega_{21}\cap\Omega_{12}$. The sample points are given by the following barycentric coordinates ($r_1,r_2,r_{20}$):  
$A=(0.8,0.1,0.1)$, 
$B=(0.33,0.33,0.33)$, 
$C=(0.62,0.07,0.31),$
 $D=(0.55,0.22,0.23),$ 
 $E=(0.52, 0.13,0.35)$,  and $F=(0.65, 0.16,0.19)$. The dashed lines indicate the unstable branches and the vertical lines indicate the predicted positions of the discontinuous phase transitions.}
\label{fig.combi3}
\end{center}
\end{figure*}

\begin{figure*}
\begin{center}
\includegraphics[width=0.9\textwidth]{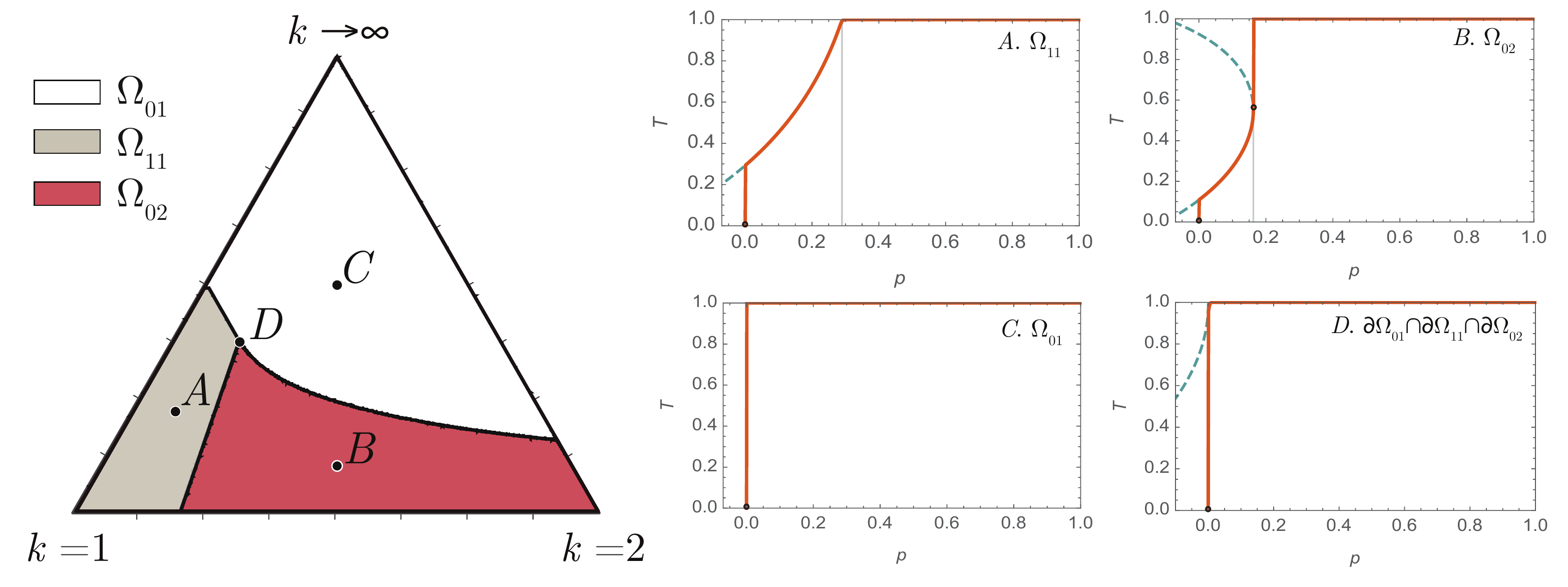}
\caption{
The barycentric plot characterizing the phase diagram of link percolation for the branching simplicial complex with $m=3$ and $r_k=r_1\delta_{k,1}+r_2\delta_{k,2}+r_{\infty}\delta_{k,\hat{k}}$ with $\hat{k}\to \infty$.
 The percolation probability $T$ versus $p$ is shown at  points A, B, and C that belong to phases $\Omega_{11}, \Omega_{02}$, and $\Omega_{01}$  respectively, and at their shared accumulation point  $D \in\Omega_{11}\cap\Omega_{01}\cap\Omega_{02}$. The sample points  are given by their barycentric coordinates ($r_1,r_2,r_{\infty}$):  
 $A=( 0.70, 0.08,0.22 )$,
 $B=(0.45,0.45,0.1)$,
 $ C=(0.25, 0.25, 0.50)$,  and $D=(0.50, 0.13, 0.37)$.
  The dashed lines indicate the unstable branches and the vertical lines indicate the predicted positions of the discontinuous phase transitions.}
\label{fig.combi4}
\end{center}
\end{figure*}

Here we provide evidence  that  random branching cell complexes can feature more percolation transitions in addition to the known upper and lower ones.  These transitions can occur for $p=p_c^{\star}$ with $p^{\star}<p_c^{\star}<p_c$ and they are characterized by discontinuities both in the percolation probability $T$ and in the fractal exponent $\psi$, with $T<1$ and $\psi<1$ above and below the  transition. In other words, the maximum cluster still remains sub-extensive after undergoing a discontinuous transition; see, for example, Fig.\ $\ref{TPsimultiple}$.  These phase transitions correspond to hybrid critical points of Eq.\ (\ref{FT}) different from the upper or lower percolation threshold and correspond to the multiple percolation transitions observed in the correlated multiplex network considered in Ref.\ \cite{kryven2019sharp} for the modified branching network model.
Here we investigate   this interesting behavior in the context of a  simplicial complex ($m=3$) with a trimodal distribution $r_k$ given by 
\bea\label{eq:example}
r_k=r_1\delta_{k,1}+r_2\delta_{k,2}+\hat{r}\delta_{k,\hat{k}},
\eea
where $r_1+r_2+r_{\hat{k}}=1$ and $\hat{k}\geq3$.
By numerically studying the roots of Eq.~(\ref{FT}) supplied with the trimodal distribution \eqref{eq:example}, we build the phase diagram of the model as a barycentric plot for various values of $\hat{k}$ (see Fig.\ \ref{fig.combi2}).
In   Fig.\ \ref{fig.combi2} we distinguish between different phases 
 $\Omega_{ij}$ corresponding to parameter values for which the percolation probability displays  $i=0,1,2$ continuous, and $j=0,1,2$ discontinuous and hybrid critical points, so that $i+j$ indicates the total number of distinct percolation thresholds.  As shown in Fig~\ref{fig.combi2}  the phase diagram evolves when $\hat{k}$ increases. Fig.~\ref{fig.combi3} gives several examples of the percolation probability $T$ as a function of $p$  in the different phases and demonstrates the existence of intermediate percolation transitions.
When  ${k}\to\infty$,  the phase diagram degenerates, and the phase diagram consists of three phases; see Fig.~\ref{fig.combi4}.
In this limit we observe a phase $\Omega_{01}$ with a discontinuous 0-to-1 transition for the percolation probability $T$ at  $p^{\star}=p_c=0$.
Interestingly by having a multi-modal $r_k$ distribution and  a random multi-modal distribution $q_m$ of the number of sides of the polygons (with more than three modes) it is possible to observe even more than one intermediate phase transition (see for instance corresponding phenomenology in correlated multiplex networks described in Ref.\ \cite{kryven2019sharp}).

\section{Generating function}
\label{gf1}
Here we investigate the properties   of the  generating functions  $\hat{T}_n(x)$ and $\hat{S}_n(x,y)$ that will be essential to characterize the different phases of percolation in the branching simplicial and cell complexes under investigation.
The  function $\hat{T}_n(x)$ is the generating function of the  
number of nodes in the connected component linked to both initial nodes of the considered random branching network. The function $\hat{S}_n(x,y)$  is the generating function for the sizes of the two connected components linked  exclusively to one of the two initial nodes of the same network. 
These generating functions are given by 
\bea
\hat{T}_n(x)&=&\sum_{\ell=0}^{\infty}t_n(\ell) x^{\ell},\nonumber \\
\hat{S}_n(x,y)&=&\sum_{\ell=0}^{\infty}\sum_{\bar{\ell}=0}^{\infty}s_n(\ell,\bar{\ell})x^{\ell}y^{\bar{\ell}},
\eea
where  $t_n(\ell)$ indicates  the distribution of the number of nodes $\ell$ connected to the two initial nodes and  $s_n(\ell,\bar{\ell})$  indicates  the joint distribution of  the number of nodes $\ell$ connected exclusively to a given initial node and the number of nodes $\bar{\ell}$ connected exclusively to the other initial node.
 By being guided by the diagrammatic representation of these quantities, as explained  in Refs.~\cite{hyperbolic_Ziff,kryven2019renormalization}, we obtain the recursive equations for $\hat{T}_n(x)$ and $\hat{S}_n(x,y)$ that  start from the initial condition $T_{0}(x)=1-\hat{S}_{0}(x,y) =p$  and read 

\begin{widetext}
\bea
\hat{S}_{n+1}(x,y)&=&(1-p)\sum_{k=1}^{\infty}r_k\left[\sum_{r=0}^{m-2}x^ry^{m-2-r}\hat{T}_n^{r}(x)\hat{S}_n(x,y)\hat{T}_n^{m-2-r}(y)+\sum_{s=0}^{m-3}\sum_{r=0}^{s}x^{r}y^{s-r}\hat{T}_n^{r}(x)\hat{S}_n(x,1)\hat{S}_n(y,1)\hat{T}_n^{s-r}(y)\right]^k.\nonumber \\
\hat{T}_{n+1}(x)&=&\sum_{k=1}^{\infty} r_k\left[x^{m-2}\hat{T}_n^{m-1}(x)+(m-1)x^{m-2}\hat{T}_n^{m-2}(x)S_n(x,x)+\sum_{s=0}^{m-3}(s+1)x^s\hat{T}_n^{s}(x)\hat{S}_n(x,1)\hat{S}_n(1,x)\right]^k\nonumber \\
&&-(1-p)\sum_{k=1}^{\infty} r_k \left[(m-1)x^{m-2}\hat{T}_n^{m-2}(x)S_n(x,x)+\sum_{s=0}^{m-3}(s+1)x^s\hat{T}_n^{s}(x)\hat{S}_n(x,1)\hat{S}_n(1,x)\right]^k,\label{TS} 
\eea
\end{widetext}

\section{Fractal exponent}
\subsection{General framework}
\label{fractal_general1}
The total number of nodes  $M_n$  that at iteration $n$ are in  the   component connected to the two initial nodes   
 can be obtained by differentiating the  generating function $\hat{T}_n(x)$ i.e.,
\bea
M_n=\left.\frac{d\hat{T}_n(x)}{dx}\right|_{x=1}.
\eea
By following the mathematical framework proposed in Ref.~ \cite{hyperbolic_Ziff} we rewrite Eqs.\ $(\ref{TS})$ in the vector  form
\bea
\mathbf V_n(x)&=& \left(V_n^1(x),V_n^2(x),V_n^3(x)\right)^\top\nonumber \\ &=&\left(\hat{T}_n(x),\Sigma_n(x),S_n(x)\right)^\top,\eea where $\Sigma_n(x)=\hat{S}_n(x,x)$, and $S_n(x)=\hat{S}(1,x)$ and obtain the recursive equation 
\bea\label{eq:poly:V}
\mathbf V_{n+1}(x) =  \mathbf F_n(\mathbf V_{n}(x), x ).
\eea
These equations are differentiated to obtain
\bea
\frac{d{\bf V}_{n+1}(x)}{dx}=\sum_{j=0}^n\sum_{s=1}^3\frac{\partial{\bf F}_j}{\partial{V}_j^{s}(x)}\frac{d{ V}_j^{s}(x)}{dx}+\frac{\partial{\bf F}_n}{\partial x},
\label{Vprime}
\eea
with initial condition ${\bf V}^{\prime}_0=(0,0,0)$ (where we do not count the initial nodes).

We note that   the non-homogeneous term ${\partial{\bf F}_n}/{\partial x}$  is subleading with respect to the homogeneous one. Therefore   for $n\gg 1$ and $T<1$,  we can express $M_n$ as
\bea
 M_{n+1} \simeq {\mathcal D}_n \prod_{n'=1}^n\lambda_{n'}{\bf u}_n,
\eea
where $\lambda_n$ and ${\bf u}_n$ are the largest eigenvalue and  the  corresponding eigenvector of the Jacobian matrix  ${\mathbf J}_n$ given by
\bea
\left[J_n\right]_{ij}=\left.\frac{\partial{F}^{i}(x)}{\partial{V}^{j}(x)}\right|_{{\bf V}(x)={\bf V}_{n}(1); x=1},
\eea
and ${\mathcal D}_n$ is given by 
\bea
{\mathcal D}_n= \left(\prod_{n'=2}^n \braket{{\bf u}_{n'}|{\bf u}_{n'-1}} \right)\braket{{\bf u}_1|\dot{\bf V}_0},
\eea
with  $\dot{\bf V}_0={\partial{\bf F}_0}/{\partial x}$.
We will show  that for $p\simeq p_c$, ${\mathcal D}_n$ is in first approximation independent of $n$, therefore it  follows that $R_n=\dot{V}_n^1$ scales like
\bea
M_{n+1}\sim \prod_{n'=1}^n\lambda_{n'}=\exp\left[{\sum_{n'=0}^n \ln \lambda_n}\right].
\label{Rna}
\eea

By using Eq.\ (\ref{Mn}) it follows that  $\psi_n$ is given by  
\bea
\psi_n=\frac{\ln \lambda_n}{\ln [\avg{k}(m-1)]}
\label{psi_n}
\eea
and the fractal exponent $\psi$ can be calculated by performing the limit for $n\to \infty$, and using the definition of the fractal exponent given by Eq.\ (\ref{psi_def}).

\subsection{Derivation of the fractal exponent $\psi$}
In this section our goal is to derive the explicit expression for the fractal exponent $\psi$. From the explicit expression of the generating functions given by  Eq.~\eqref{TS}, we derive the closed equations for $\mathbf V_n=[\hat{T}_n(x),\Sigma_n(x),S_n(x)]^\top$, where $\Sigma_n(x)=\hat{S}_n(x,x)$, and $\hat{S}_n(x)=\hat{S}(1,x)$
\bea\label{eq:sim:V}
\mathbf V_{n+1}(x) =  \mathbf F(\mathbf V_{n}(x), x ).
\eea
These equations read
\begin{widetext}
\bea
\hat{T}_{n+1}(x) &=&\sum\limits_{k=1}^\infty r_k \left[x^{m-2}\hat{T}_n^{m-1}(x)+ (m-1)  x^{m-2}   \hat{T}^{m-2}_n(x)\Sigma_n(x)+ \left(\sum_{i=0}^{m-3}(i+1) x^i \hat{T}_n^i(x) \right)S_n^2(x)\right]^k\nonumber \\
 &&-(1-p)\sum\limits_{k=1}^\infty r_k\left[(m-1)x^{m-2} \hat{T}_n^{m-2}(x) \Sigma_n(x) +\left(\sum_{i=0}^{m-3}(i+1)x^i \hat{T}_n^i(x)\right) S^2_n(x) \right]^k,\nonumber \\
\Sigma_{n+1}(x) &=&(1-p)\sum\limits_{k=1}^\infty r_k\left[(m-1)x^{m-2} \hat{T}_n^{m-2}(x) \Sigma_n(x) +\left(\sum_{i=0}^{m-3}(i+1)x^i \hat{T}_n^i(x)\right) S^2_n(x) \right]^k,\nonumber \\
S_{n+1}(x) &=& (1-p)\sum\limits_{k=1}^\infty r_k\left[\left(\sum_{i=0}^{m-2}x^i\hat{T}_n^i(x)\right)S_n(x)\right]^k.
\label{TS2}
\eea
\end{widetext}
The Jacobian $\mathbf{J}_n$ is obtained by differentiating Eq.\ \eqref{TS2} with respect to $\mathbf{V}_n$, by putting $x=1$ and using $\hat{T}_n(1)=1-\Sigma_n(1,1)=1-S_n(1)=T_n$.
In order to perform this analytical calculation we have used the  mathematical identities
\bea
(1-T_n)\sum_{i=0}^{m-3} i(i+1)T_n^{i-1}&=&2\sum_{i=0}^{m-3}(i+1)T_n^i\nonumber \\
&&-(m-1)(m-2)T_n^{m-3}\nonumber \\
\eea
and
\bea
(1-T_n)\sum_{i=0}^{m-3}(i+1)T_n^i&=&\sum_{i=0}^{m-2}T_n^i-(m-1)T_n^{m-2}.\nonumber
\eea
In this way it is easy to show  that the Jacobian ${\mathbf J}_n$ can be expressed as
\begin{widetext}
\bea
\hspace*{-7mm}{\mathbf J}_n=\left(
\begin{array}{ccc}
 \avg{k}[2H(T_n)-Q^{\prime}(T_n)]-2(1-p) G(T_n)&  Q'(T_n)[\avg{k}-(1-p)R^{\prime}(1-T_n^{m-1}] & 2 \left[\avg{k}(H(T_n)-Q^{\prime}(T_n))-(1-p)G(T_n)\right] \\
 2 (1-p)G(T_n)& (1-p) R^{\prime}(1-T_n^{m-1})Q'(T_n) & 2 (1-p) G(T_n) \\
 (1-p) G(T_n) & 0 &(1-p)R^{\prime}(1-T_n^{m-1})\text{H}(T_n) \\
\end{array}
\right),\nonumber
\eea
\end{widetext}
where  
$Q(T)$  is defined in Eq.\ (\ref{QT}) and $H(T)$ is defined as
\bea
H(T)=\sum_{i=0}^{m-2} T^i,
\label{H}
\eea
which admits for  $T<1$ the expression
\bea
H(T)=\frac{1-Q(T)}{1-T}.
\label{H2}
\eea
Furthermore, 
$R^{\prime}(z)$ and $G(T)$ are given by  
\bea
R^{\prime}(z)&=&\sum_k kr_k z^{k-1},
\nonumber \\
G(T)&=&R^{\prime}(1-T_n^{m-1})\left[\text{H}(T_n)-Q'(T_n)\right].
\eea
Using the mathematical identities listed above and using a procedure similar to the one used for deriving the expression of the Jacobian,  it can be shown that ${\partial{\bf F}_n}/{\partial x}$ is given by 
\begin{widetext}
\bea
\frac{\partial{\bf F}_n}{\partial x}=\left(\begin{array}{c} \avg{k}T_n2[H(T_n)-Q^{\prime}(T_n)]+\avg{k}[T_nQ^{\prime}(T_n)-Q(T_n)]-2(1-p)T_nG(T_n)\\2(1-p)T_nG(T_n)\\(1-p)T_nG(T_n)\end{array}\right).\nonumber
\eea
\end{widetext}
For  $T_n<1$  the Jacobian ${\mathbf{J}}_n$ has the largest eigenvalue $\lambda_n$ given by
\bea
\lambda_n&=&\frac{1}{2} \left[\sqrt{\hat{\Delta}(T_n)}+ \hat{K}(T_n) \right],
\label{lambda_n}
\eea
where $\hat{\Delta}(T_n)$ and $\hat{K}(T_n)$ are  given by 
\bea
\hspace*{-5mm}\hat{\Delta}(T_n)&=&\left[ \hat{K}(T_n)\right]^2-4  (1-{p})H(T_n) Q'(T_n)R^{\prime}(1-T_n^{m-1})\avg{k},\nonumber \\
\hspace*{-5mm}\hat{K}(T_n)&=&[2\avg{k}-(1-p)R^{\prime}(1-T_n^{m-1})]H(T_n)\nonumber \\
\hspace*{-5mm}&&- [\avg{k}-2(1-p)R^{\prime}(1-T_n^{m-1})]Q'(T_n).
\eea
For  $T_n=1$, instead, the largest eigenvalue is given by
{$$\lambda_n=\Avg{k}(m-1).$$
The eigenvector ${\bf u}_n$ corresponding to the largest eigenvalue is 
\bea
\hspace*{-7mm}{\bf u}_n={\mathcal C}\left(\begin{array}{c}\hat{K}(T_n)-2(1-p)R^{\prime}(1-T_n^{m-1})H(T_n)+\sqrt{\hat{\Delta}(T_n)}\\
4(1-p)R^{\prime}(1-T_n^{m-1})[H(T_n)-Q^{\prime}(T_n)] \\ 
2(1-p)R^{\prime}(1-T_n^{m-1})[H(T_n)-Q^{\prime}(T_n)]
\end{array}\right),\nonumber
\eea
where ${\mathcal C}$ is the normalization constant.
For  $T_n=1$ and $p=p_c$,  the  eigenvector ${\bf u}_n$ is given by
$${\bf u}_n=\left(1,0,0\right)^{\top}.$$}
Finally by using Eqs.\ $(\ref{psi_def})$ and  $(\ref{psi_n})$ we can determine  the fractal exponent $\psi$ starting from the explicit expression of the eigenvalue $\lambda_n$ given by Eq.\ (\ref{lambda_n}).

\subsection{Critical scaling of the fractal exponent}

\begin{figure}
\begin{center}
\includegraphics[width=0.9\columnwidth]{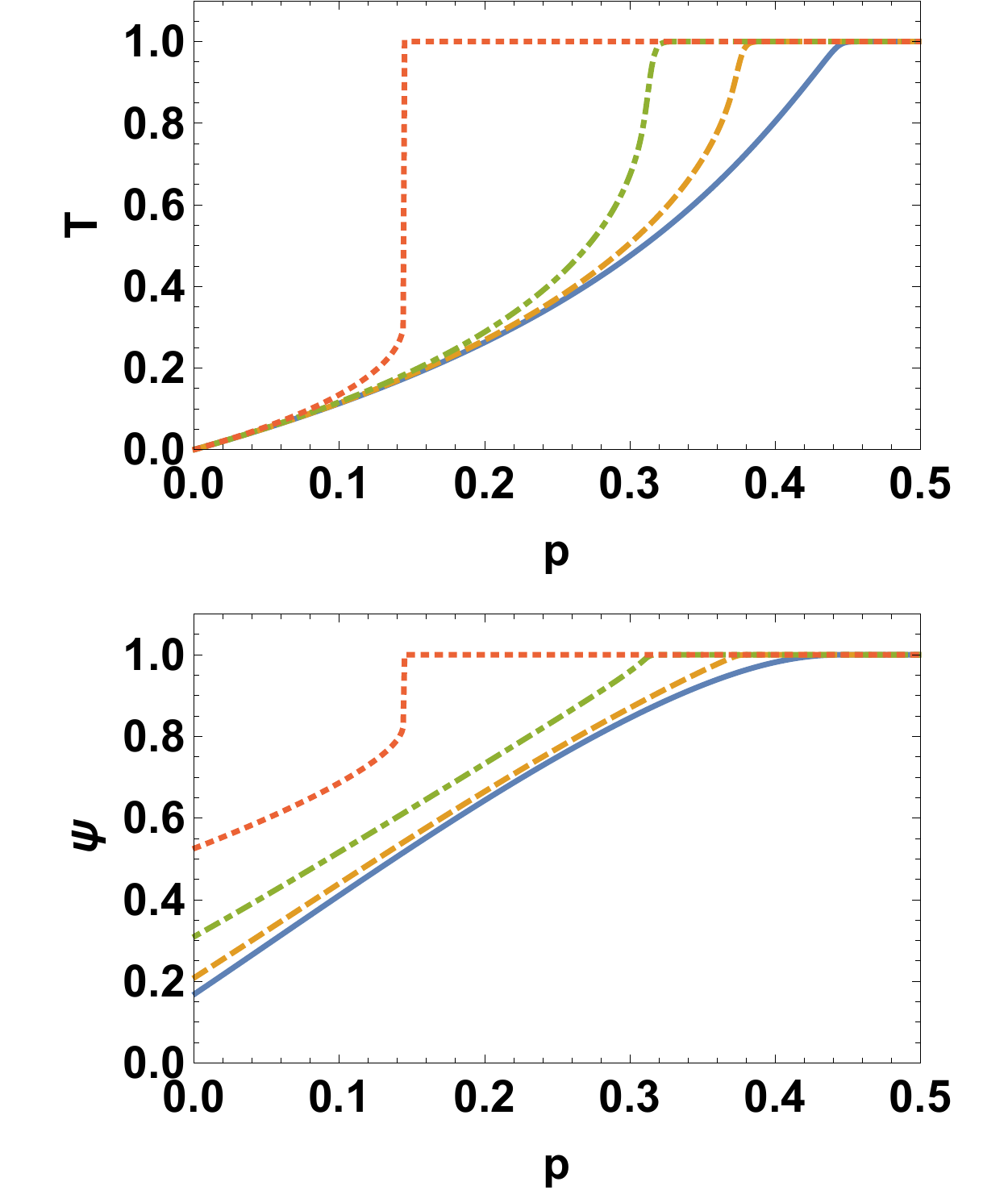}
\caption{ The percolation probability $T$ and the fractional exponent $\psi$ are plotted as a function of  occupation probability $p$  of the links. Here we consider a branching simplicial complex $(m=3)$ with $r_k$ distribution $r_k=r_1\delta_{k,1}+r_2\delta_{k,2}+r_3\delta_{k,3}$ with $(r_1,r_2,r_3)$ given by $(0.9,0.05,0.05)$ for the solid blue line (continuous  critical point of Eq.\ (\ref{FT})), $(0.8,0.2,0)$ for the orange dashed line (tricritical point of Eq.\ (\ref{FT})), $(0.727273,0.181818,0.0909091)$ for the green dot-dashed line ($s$ order critical point of Eq.\ (\ref{FT}) with $s=4$) and $(0.15,0.54,0.31)$ for the red dotted line (discontinuous  critical point of Eq.\ (\ref{FT})).}
\label{fig.TPSIstd}
\end{center}
\end{figure}

Here we consider the critical scaling of the fractal exponent close to the upper percolation threshold  in the case in which the critical percolation probability $T_c=1$ is reached continuously by the solution of Eq.\ (\ref{FT}).
When $T_c=1$, by expanding $Q'(T_n)$ and $H(T_n)$ close to the critical point, i.\,e., for  $p=p_c+\Delta p$,  and  $T_n=T_c+\Delta T_n$  for $\Delta p<0$ and $\Delta T_n<0$ but small in absolute values, i.\,e., $|\Delta T_n|\ll1$ and $|\Delta p|\ll1$, we obtain
\bea
Q'(T_n)&=&({m}-1)+{(m-1)(m-2)}\Delta T_n+\nonumber \\&&\hspace{-7mm}\frac{1}{2}{(m-1)(m-2)(m-3)}(\Delta T_n)^2+o((\Delta T_n)^2),\nonumber \\
H(T_n)&=&({m}-1)+\frac{1}{2}{(m-1)(m-2)}\Delta T_n\nonumber \\&&\hspace{-7mm}+\frac{1}{6}{(m-1)(m-2)(m-3)}(\Delta T_n)^2+o((\Delta T_n)^2).\nonumber
\eea
Moreover we can also expand the expression $R^{\prime}(1-T_n^{m-1})$ obtaining
\bea
&&R^{\prime}(1-T_n^{m-1})\simeq r_1-2r_2(m-1) \Delta T_n\nonumber \\
&&+[-r_2(m-1)(m-2) +3r_3(m-1)^2](\Delta T_n)^2+o((\Delta T_n)^2).\nonumber
\eea
By using the definition of $\psi_n$ (Eq.\ (\ref{psi_n})) and the explicit expression of $\lambda_n$ (Eq.\ (\ref{lambda_n})) we can derive the scaling of $\psi_n$ as a function of $\Delta T_n$ 
\bea
\psi_n\simeq 1-a (\Delta T_n)^2
\label{crit_psi}
\eea
where   $a$ is a constant given by 
\bea
a&=&\frac{(m-2)}{[6\avg{k}^2(m-1)-6]\ln[\avg{k}(m-1)]}\nonumber \\
&&\times\left[\avg{k}^2(m-3)(m-1)+2m-3\right]\nonumber
\eea
Therefore, for the branching cell complexes considered in this work, as long at the critical percolation probability $T_c=1$ is reached continuously by the solution of  Eq.\ (\ref{T}), the critical scaling of $\psi_n$ is universally dictated by Eq.\ (\ref{crit_psi}) (note  however that this critical behavior can be altered if the size of the polygons $m$ is randomly distributed and its distribution is fat-tailed \cite{kryven2019renormalization}).
From the universal scaling of $\psi_n$ as a function of $\Delta T_n$  we can derive the critical behavior of the fractal exponent as a function of $\Delta T$ by performing the limit $n\to \infty$. In Sec.\ III we have shown that the  scaling of $\Delta T$ with $\Delta p$ can be characterized by any exponent of the type $\beta=1/(s-1)$ when $T_c=1$ is reached continuously. This implies that  the fractal exponent scales like
\bea
\psi\simeq 1-\tilde{a} (\Delta p)^{2\beta},
\eea
where $\tilde{a}$ is a constant.
For the topologies considered in this work the only possible deviation  from the universal critical scaling $\psi_n$ as a function of $\Delta T_n$ given by Eq.\ (\ref{crit_psi}) is observed when at the upper percolation threshold the percolation probability is discontinuous. In this case we will also observe a discontinuity of the fractal exponent $\psi$ at  $p_c$. In Fig.\ \ref{fig.TPSIstd} we display the percolation probability $T$ as a function of $p$ for branching cell complexes undergoing percolation transitions of different universality classes at the upper percolation threshold $p=p_c$.

\section{Order parameter}
\subsection{General framework}
In this section we use   the  RG technique  \cite{RG,kryven2019renormalization} to predict the nature of the percolation phase transition at the upper critical percolation threshold $p_c$. At $p_c$ the order   parameter is given by the fraction $P_{\infty}$ of nodes in the giant component in an infinite network given by Eq.\ (\ref{Pinfty0}), which we rewrite here for convenience,
\bea
P_{\infty}&=&\lim_{n\to \infty}\frac{M_n}{\bar{N}_n}.
\eea
By using Eq.\ (\ref{Rna}) for approximating $M_n$ when $n\gg1$ we obtain
\bea
P_{\infty}&\simeq&\lim_{n\to \infty}\frac{1}{\bar{N}^{(0)}_n}\prod_{n'=1}^{n}\lambda_{n'}\nonumber \\
&\simeq &\exp\left[-\ln[\avg{k}({m-1})]\int_0^{\infty}dn(1-\psi_n)\right].
\label{pinft}
\eea
The RG flow  can be derived directly by the  RG  Eq.\ (\ref{RG}), which we rewrite here for convenience as
\bea
T_{n+1}={F}(p,T_n)=1-(1-p)\sum_{k}r_k (1-T_n^{m-1})^k.
\label{RG2}
\eea
In the RG procedure one proceeds as follows.

\begin{figure}
\begin{center}
\includegraphics[width=0.9\columnwidth]{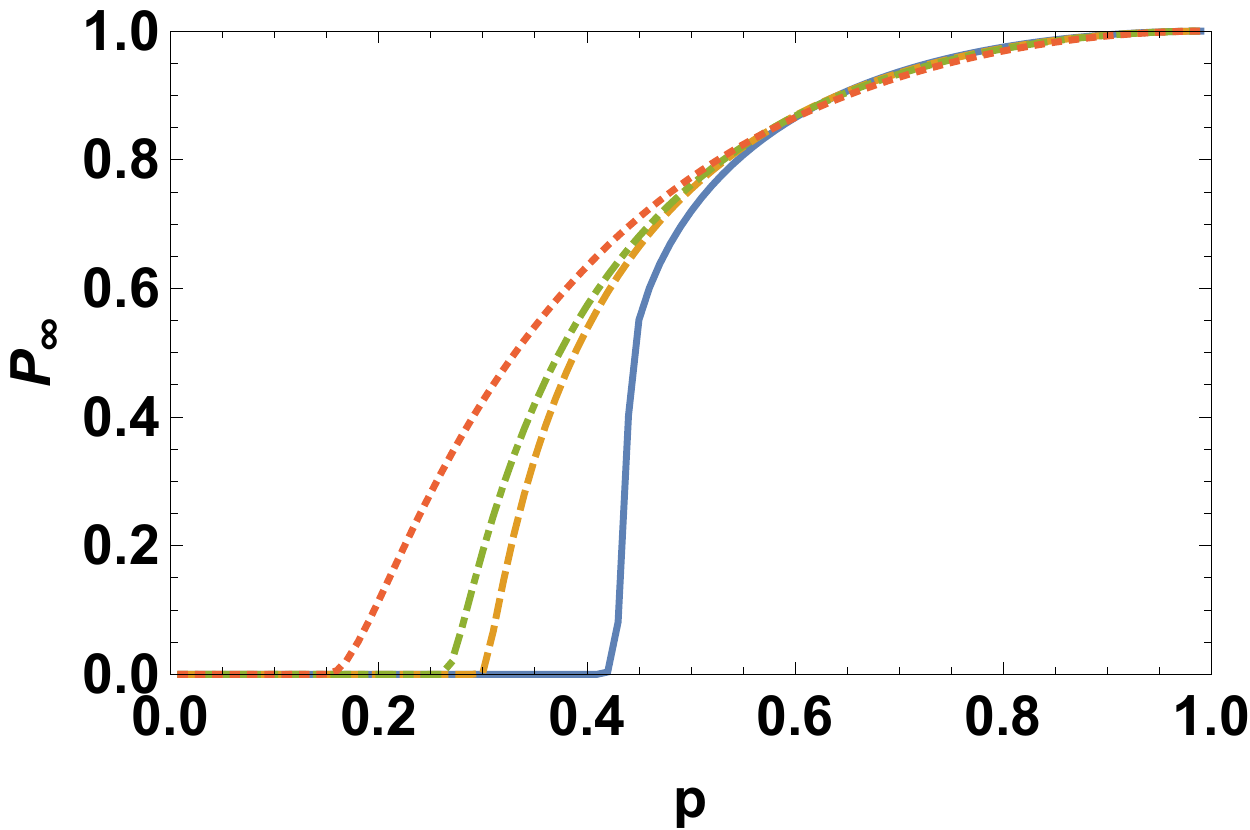}
\caption{The percolation order parameter $P_{\infty}$ as a function of  occupation probability $p$  of the links. Here we consider a branching simplicial complex $(m=3)$ with $r_k$ distribution $r_k=r_1\delta_{k,1}+r_2\delta_{k,2}+r_3\delta_{k,3}$ with $(r_1,r_2,r_3)$ given by $(0.9,0.05,0.05)$ for the solid blue line (discontinuous percolation transition),    $(0.8,0.2,0)$ for the orange dashed line (second-order transition), $(0.727273,0.181818,0.0909091)$ for the green dot-dashed line (anomalous transition following Eq.\ $(\ref{anomalous})$ with $\sigma=3$) and $(0.15,0.54,0.31)$ for the red dotted line (BKT transition). The figure has been obtained by iterating  Eq.\ (\ref{Vprime}) for $n=1000$ iterations.}
\label{fig.pinf3}
\end{center}
\end{figure}

First, the RG equation (\ref{RG2}) is expanded  close to   the critical point $(p,T)=(p_c,T_c)$,  obtaining the scaling  of $\Delta T_n=T_n-T_c$ with $n$, for  $0<\Delta p=p-p_c\ll 1$.
Secondly, this scaling is inserted  in  Eq.\ (\ref{crit_psi}) characterzing the critical behavior of $1-\psi_n$ as a function of $\Delta T_n$. 
Finally,  using Eq.\ (\ref{pinft}), we can predict the nature of the phase transition by deriving the scaling of the order parameter $P_{\infty}$ close to the upper percolation threshold. 
Here we conduct this  RG study in the different phases of percolation defined on branching cell complexes and we explain the different critical behavior that can be observed for the percolation order parameter $P_{\infty}$ (see Fig.\ \ref{fig.pinf3}).

In the following sections we will use the RG technique to predict  that at the transcritical bifurcation point the percolation transition is discontinuous, similar to the Farey graph \cite{hyperbolic_Ziff} and well-behaved generalized  $2$d  hyperbolic manifolds \cite{kryven2019renormalization}; at the saddle-node bifurcation point we predict a BKT transition and at the third-order pitchfork bifurcation point we predict a second-order critical behavior (see Fig.\ \ref{fig.pinf3}). These results confirm and generalize previous results obtained in hierarchical networks and specifically the flower network \cite{Patchy,flower_tau,tricritical,RG}. Here we reveal additional universality classes that can occur at the pitchfork singularities of order $s>3$ where we predict and observe a  critical scaling of the type 
\bea
P_{\infty}\simeq \exp\left[-{\mathcal A}/(\Delta p)^{\sigma}\right]
\label{anomalous}
\eea 
with the theoretically derived anomalous exponent $\sigma=(s-3)/(s-1)$.

\subsection{RG theory at the transcritical bifurcation point}

At the transcritical bifurcation point we observe a discontinuous percolation transition that is in the same universality class as percolation in the Farey graph \cite{hyperbolic_Ziff} and in well-behaved $2d$ cell complexes \cite{kryven2019renormalization}.
The derivation of this result  follows   steps similar to the ones previously reported in Ref.~\cite{kryven2019renormalization}. Here we report  these results for the self-consistency of this work.

Our goal is to study the critical behavior above  the upper percolation threshold, for $\Delta p=p-p_c>0$.  We  expand the RG equation $(\ref{RG2})$ for $0<\Delta p\ll1$  and $|\Delta T_n|=|T-T_c|\ll1$ when  $p_c=1-{1}/{[(m-1)r_1]}$ and $T_c=1$ obtaining
\bea
T_{n+1}&=&{F}(p_c,T_c)+\left.\frac{\partial {F}}{\partial p}\right|_{p=p_c,T=T_c}\Delta p\nonumber \\
&&+\left.\frac{\partial F}{\partial T}\right|_{p=p_c,T=T_c}\Delta T_n+\left.\frac{\partial^2 F}{\partial p\partial T}\right|_{p=p_c,T=T_c}\Delta p \Delta T_n\nonumber \\
&&+\frac{1}{2}\left.\frac{\partial^2 F}{\partial T^2}\right|_{p=p_c,T=T_c}(\Delta T_n)^2+\ldots
\eea
with
\bea
F(p_c,T_c)&=&T_c=1,\nonumber \\
\left.\frac{\partial F}{\partial p}\right|_{p=p_c,T=T_c}&=&0,\\
\left.\frac{\partial F}{\partial T}\right|_{p=p_c,T=T_c}&=&1,\\
\left.\frac{\partial^2 F}{\partial p\partial T}\right|_{p=p_c,T=T_c}&=&-r_1({m-1}),\\
\left.\frac{\partial^2 F}{\partial T^2}\right|_{p=p_c,T=T_c}&=&\frac{{r_1}(m-2)-2r_2(m-1)}{r_1}\\
\eea

Therefore by truncating the expansion to the leading terms in $\Delta T_n$ and $\Delta p$ we can write
\bea
\Delta T_{n+1}-\Delta T_n=\hat{C}\Delta T_n \left[\Delta T_n-\hat{B}\Delta p\right],
\label{uno}
\eea
with  constants $\hat{B}$ and $\hat{C}$ given by
\bea
\hat{B}=\frac{2(m-1)r_1^2}{(m-2)r_1-2{r_2}(m-1)},\\
\hat{C}=\frac{1}{2r_1}\left[(m-2)r_1-2{r_2}(m-1)\right].
\eea
For  $n\to \infty$, we adopt a  continuous approximation of Eq.\ (\ref{uno}). We   indicate with   $x$  the continuous approximation of $-\Delta T_n\ll 1 $, i.e.,  $x\simeq -\Delta T_n$, that follows the  differential equation 
\bea
\frac{dx}{dn}=-\hat{C}x[x+\hat{B} \Delta p ],
\label{xn}
\eea
with initial condition $x(0)=1-p$. This equations has the solution  
\bea
x(n)={\hat{B}\Delta p}\left[ \left(1+\frac{\hat{B}\Delta p}{1-p}\right)e^{\hat{C}\hat{B}\Delta p n}-1\right]^{-1}.
\label{xn1}
\eea

For $r_1>2r_2(m-1)/(m-2)$, $\psi_n$ obeys the scaling relation Eq.\ (\ref{crit_psi}) that can be expressed as a function of $x(n)$ as
\bea
\psi_n=1-a(T_c-T_n)^2=1-a [x(n)]^2.
\eea
Consequently, using Eq.\ (\ref{pinft}) we can express $P_{\infty}$ in the continuous approximation as
\bea
P_{\infty} (p)&\simeq &\exp\left[-\ln[\avg{k} (m-1)] a\int dn [x(n)]^2\right].
\eea
Using the expression of $x(n)$ given by Eq.\ (\ref{xn1}) we obtain for $0<p-p_c\ll 1$
\bea
P_{\infty} (p)&\simeq
&\exp\left[-\ln[\avg{k}(m-1)]a\left(\frac{(1-p)}{\hat{C}}\right.\right.\nonumber \\
&&\left.\left.+\frac{\hat{B}\Delta p}{\hat{C}}\ln\left(\frac{\hat{B}\Delta p}{\hat{C}}\right)\right)\right]\eea
which can  also be written as 
\bea
P_{\infty} (p)&\simeq &P_{\infty}(p_c)\left(\frac{\Delta p}{r}\right)^{-h \Delta p}\nonumber \\
\label{Pd1}
\eea
where $P_{\infty}(p_c), h$ and $r$ are given by  
\bea
P_{\infty}(p_c)&=&\exp\left[-\ln[\avg{k}({m}-1)]a\frac{2}{{(m-1)(m-2)}}\right],\nonumber\\
h&=&\ln[\avg{k}({m}-1)]a\frac{\hat{B}}{\hat{C}},\nonumber \\
r&=&\frac{\hat{C}}{\hat{B}({m-1})}.
\eea
Eq.\ (\ref{Pd1}) can be further expanded for $0<\Delta p\ll1 $ obtaining 
the critical behavior 
\bea
P_{\infty} (p)&\simeq &P_{\infty}(p_c)+\alpha \Delta p \left[-\ln \left(\Delta p\right)\right],
\eea
where $\alpha=P_{\infty}(p_c)h.$

\subsection{RG theory at the saddle-node bifurcation point}
In this section we follow Ref.\ \cite{tricritical,RG} and  by using the RG theory we show that as long as $T_c<1$ the upper percolation threshold follows a BKT transition.
Developing Eq.\ (\ref{RG2}) for $T_n=T_c+\Delta T_n$ and $p=p_c+\Delta p$ with  $0<\Delta p\ll 1$ up to  second order we obtain
\bea
T_{n+1}&\simeq &F(p_c,T_c)+\left.\frac{\partial F}{\partial p}\right|_{p=p_c,T=T_c}\Delta p\nonumber \\&&+\left.\frac{\partial F}{\partial T}\right|_{p=p_c,T=T_c}\Delta T_n
+\frac{1}{2}\left.\frac{\partial^2 F}{\partial T^2}\right|_{p=p_c,T=T_c}(\Delta T_n)^2,\nonumber
\eea
with
\bea
F(p_c,T_c)&=&T_c<1,\nonumber \\
\left.\frac{\partial F}{\partial p}\right|_{p=p_c,T=T_c}&=&a=\frac{1-T_c}{1-p_c}>0,\\
\left.\frac{\partial F}{\partial T}\right|_{p=p_c,T=T_c}&=&1,\\
\left.\frac{\partial^2 F}{\partial T^2}\right|_{p=p_c,T=T_c}&=&2b>0.\\
\eea
Therefore close to the upper percolation transition $\Delta T_n$ evolves according to the equation 
\bea
\Delta T_{n+1}-\Delta T_n=a\Delta p+b(\Delta T_n)^2.
\eea
For $|\Delta T_n|\ll1$ we can write the above equation in the continuous limit as 
\bea
\frac{dy(\hat{n})}{d\hat{n}}=1+y^2,
\eea
where $\hat{n}=n\delta$ and $\delta=\sqrt{ab\Delta p}$ and $y=b\Delta T_n/\delta$. This equation has  the solution
\bea
y(\hat{n})=\tan(\hat{n}+\tan^{-1}[y(0)]),
\eea
displaying a divergence for $\hat{n}_c=n_c\delta$ such that $\hat{n}_c+\tan^{-1}(y(0))=\pi/2$.
Therefore, the continuous approximation of $\Delta T_n$ indicated by $x(n)$ obeys
\bea
x(n)=\frac{\delta}{b}\tan({n}\delta+\tan^{-1}[y(0)]) . 
\eea 
The initial condition $x(0)=(T_0-T_c)\leq 0$ implies that  $y(0)\leq 0$.
However, the function $x(n)$ eventually becomes positive and diverges for  $n={n_c}=\hat{n}_c/\delta$. At $n_c\simeq \hat{n}_c/\delta$ the approximation $x(n)\ll1$ is no longer valid and the solution $x(n)$ has a jump to the trivial solution $x_0=1-T_c$.  \begin{figure}
 \begin{center}
\includegraphics[width=\columnwidth]{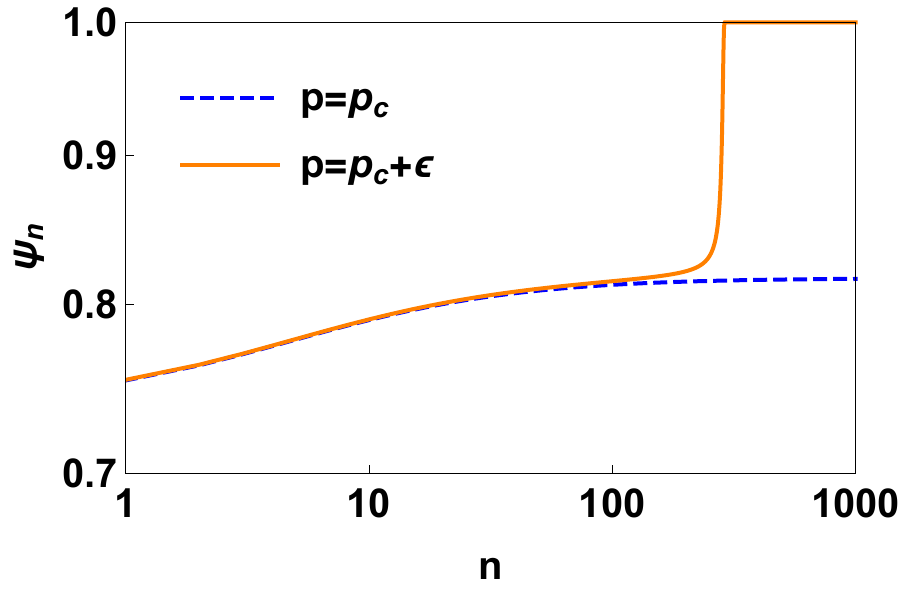}
\caption{ The scaling of the exponent $\psi_n$ as a function of $n$ is plotted at the critical point $p=p_c$ corresponding to the BKT transition and slightly above the critical point for $p=p_c+\epsilon$ with $\epsilon=10^{-4}$.  The $r_k$ distribution is given by $r_k=0.1\delta_{k,1}+0.5\delta_{k,2}+0.3\delta_{k,3}+0.1 \delta_{k,4}$   and $p_c=0.1261(7)$.}
\label{fig.psiBKT}
\end{center}
\end{figure}
In the latter case, we have that $1-\psi_n$ will also have a discontinuity  at $n_c$ (see Fig.\ \ref{fig.psiBKT}), i.e.,
\bea
1-\psi_n=\left\{\begin{array}{ccc}f_{\psi}(\hat{n})&\mbox {for} &n<n_c \\
0&\mbox{for}& n>n_c\end{array}.\right.
\eea
Therefore, the critical transition is continuous and follows the BKT singularity. In fact we have 
\bea
P_{\infty}&\simeq &\exp\left[-\int_0^{\infty} dn (1-\psi_n)\right]\nonumber \\
&\simeq &\exp\left[-\frac{1}{\delta}\int_0^{\hat{n}_c}d\hat{n} f_{\psi}(\hat{n})\right]\simeq \exp\left[-\frac{\alpha}{\sqrt{\Delta p}}\right],
\eea
where $\alpha$ is a constant.
This BKT transition has already been reported for specific branching cell complexes  and other $2$d hierarchical networks  in Refs.~\cite{Patchy,flower_tau,tricritical} and in $3$d hyperbolic manifolds in Ref.\ \cite{Bianconi_Ziff}.

\subsection{RG theory at the pitchfork bifurcation points}
In this section we study the nature of the percolation transition of the pitchfork bifurcation points. We start with the treatment of the tricritical point of Eq.\ (\ref{FT}), finding that in this case the transition is second order and subsequently we discuss the case of critical points of order $s>3$ of Eq.\ (\ref{FT}) finding continuous phase transitions with anomalous critical behavior.

Let us  expand the RG equation $(\ref{RG2})$ close to the tricritical point $(p_c,T_c)$ for $0<\Delta p\ll1$  and $|\Delta T_n|\ll1$ when $r_1=2r_2(m-1)/(m-2)$, $p_c=1-{1}/{[\Avg{m-1}r_1]}$ and $T_c=1$. In this way we obtain
\bea
T_{n+1}&
\simeq&{F}(p_c,T_c)+\left.\frac{\partial {F}}{\partial p}\right|_{p=p_c,T=T_c}\Delta p\nonumber \\
&&+\left.\frac{\partial F}{\partial T}\right|_{p=p_c,T=T_c}\Delta T_n+\left.\frac{\partial^2 F}{\partial p\partial T}\right|_{p=p_c,T=T_c}\Delta p \Delta T_n\nonumber \\
&&+\frac{1}{2}\left.\frac{\partial^2 F}{\partial T^2}\right|_{p=p_c,T=T_c}(\Delta T_n)^2\nonumber\\
&&+\frac{1}{6}\left.\frac{\partial^3 F}{\partial T^3}\right|_{p=p_c,T=T_c}(\Delta T_n)^3,
\eea
with
\bea
F(p_c,T_c)&=&T_c=1,\nonumber \\
\left.\frac{\partial F}{\partial p}\right|_{p=p_c,T=T_c}&=&0,\\
\left.\frac{\partial F}{\partial T}\right|_{p=p_c,T=T_c}&=&1,\\
\left.\frac{\partial^2 F}{\partial p\partial T}\right|_{p=p_c,T=T_c}&=&-r_1({m-1}),\\
\left.\frac{\partial^2 F}{\partial T^2}\right|_{p=p_c,T=T_c}&=&0,\\
\left.\frac{\partial^3 F}{\partial T^3}\right|_{p=p_c,T=T_c}&<&0.
\eea
Therefore, by truncating the expansion to the leading terms in $\Delta T_n$ and $\Delta p$ we can write
\bea
\Delta T_{n+1}-\Delta T_n=-\hat{C}\Delta T_n \left[(\Delta T_n)^2+\hat{B}\Delta p\right],
\eea
where constants $\hat{B}$ and $\hat{C}$ are given by
\bea
\hat{B}&=&\frac{r_1(m-1)}{\hat{C}},\\
\hat{C}&=&-\frac{1}{6}\left.\frac{\partial^3 F}{\partial T^3}\right|_{p=p_c,T=T_c}.
\eea
For  $n\to \infty$ we approximate the above equation in the continuous limit and we  use  $x$ to indicate the continuous approximation of $-\Delta T_n\ll 1 $, i.e.,  $x\simeq -\Delta T_n$. In this way we   get  the  differential equation
\bea
\frac{dx}{dn}=-\hat{C}x[x^2+\hat{B} \Delta p ],
\label{xn2}
\eea
with initial condition $x(0)=1-p$, whose solution is
\bea
x(n)=\sqrt{\hat{B}\Delta p}\left[\left(1+\frac{\hat{B}\Delta p}{(1-p)^2}\right)e^{2\hat{C}\hat{B}\Delta p n}-1\right]^{-1/2}.
\eea
Therefore the percolation order parameter $P_{\infty}$ is given by 
\bea
P_{\infty}(p)&\simeq &\exp\left[-\ln[\avg{k}(m-1)]a\int_0^{\infty}[x(n)]^2\right]\nonumber \\
&\simeq&\left(\frac{\hat{B}\Delta p}{(1-p)^2}\right)^{\hat{\beta}}
\propto (\Delta p)^{\hat{\beta}},
\eea
where 
\bea
\hat{\beta}=\ln[\avg{k}(m-1)]\frac{a}{2\hat{C}}.
\eea
It follows that in this case the transition is continuous with a critical exponent $\hat{\beta}$. This phase transition has been reported in the case of the flower network in Ref.\ \cite{tricritical}.

At the higher-order critical points of Eq.\ (\ref{FT}) of order $s$   we have \bea
F(p_c,T_c)&=&T_c=1,\nonumber \\
\left.\frac{\partial F}{\partial p}\right|_{p=p_c,T=T_c}&=&0,\\
\left.\frac{\partial F}{\partial T}\right|_{p=p_c,T=T_c}&=&1,\\
\left.\frac{\partial^2 F}{\partial p\partial T}\right|_{p=p_c,T=T_c}&=&-r_1({m-1}),\\
\left.\frac{\partial^j F}{\partial T^j}\right|_{p=p_c,T=T_c}&=&0,\\
(-1)^{s}\left.\frac{\partial^s F}{\partial T^s}\right|_{p=p_c,T=T_c}&>&0,
\eea
for $j=1,2,\ldots, (s-1)$.
Therefore  by expanding the RG equation $(\ref{RG2})$ close to the $s$-critical point $(p_c,T_c)$ for $0<\Delta p\ll1$  and $|\Delta T_n|\ll1$ when $r_1=2r_2(m-1)/(m-2)$, $p_c=1-{1}/{[\Avg{m-1}r_1]}$ and $T_c=1$ up to order $s$ we get
\bea
T_{n+1}&
\simeq&{F}(p_c,T_c)+\left.\frac{\partial {F}}{\partial p}\right|_{p=p_c,T=T_c}\Delta p\nonumber \\
&&+\left.\frac{\partial F}{\partial T}\right|_{p=p_c,T=T_c}\Delta T_n+\left.\frac{\partial^2 F}{\partial p\partial T}\right|_{p=p_c,T=T_c}\Delta p \Delta T_n\nonumber \\
&&+\frac{1}{s!}\left.\frac{\partial^s F}{\partial T^s}\right|_{p=p_c,T=T_c}(\Delta T_n)^s.
\eea

Therefore, by truncating the expansion to the leading terms in $\Delta T_n$ and $\Delta p$ we can write
\bea
\Delta T_{n+1}-\Delta T_n=-\hat{C}\Delta T_n \left[(-1)^{s-1}(\Delta T_n)^{s-1}+\hat{B}\Delta p\right],\nonumber
\eea
where constants $\hat{B}$ and $\hat{C}$ are given by
\bea
\hat{B}&=&\frac{r_1(m-1)}{\hat{C}},\\
\hat{C}&=&(-1)^s\frac{1}{s!}\left.\frac{\partial^s F}{\partial T^s}\right|_{p=p_c,T=T_c}.
\eea
By performing the limit   $n\to \infty$ we can derive the   equation  for   the continuous approximation of $-\Delta T_n\ll 1,$  indicated as $x(n)$, i.e.,  $x\simeq -\Delta T_n$. This equation reads
\bea
\frac{dx}{dn}=-\hat{C}x[x^{s-1}+\hat{B} \Delta p ],
\label{xn3}
\eea
 with initial condition $x(0)=1-p$.
  \begin{figure}
 \begin{center}
\includegraphics[width=\columnwidth]{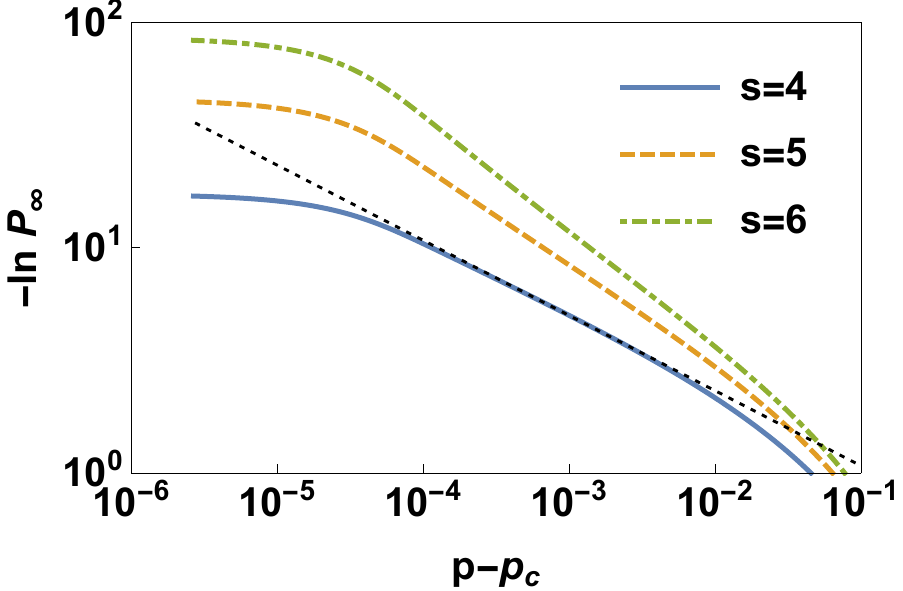}
\caption{The critical scaling of the order parameter $P_{\infty}$ versus $p-p_c$ is shown for  branching simplicial complexes ($m=3$) with distribution $r_k=r_1\delta_{k,1}+r_2\delta_{k,2}+r_3\delta_{k,3}+r_4\delta_{k,4}+r_{5}\delta_{k,5}$ and parameter values corresponding to the critical points of order $s$ of Eq.\ (\ref{FT}).   Here the data are obtained by iterating Eq.\ (\ref{Vprime}) $n=2 \cdot 10^4$ times. The thin dotted line is the theoretically predicted scaling given by Eq.\ (\ref{anomalous}) with  $\sigma=1/3$.}
\label{fig.anomalous}
\end{center}
\end{figure}
This equation has the solution 
\bea
x(n)=\left(\hat{B}\Delta p\right)^{\gamma}\left[\left(1+\frac{\hat{B}\Delta p}{(1-p)^{s-1}}\right)e^{(s-1)\hat{C}\hat{B}\Delta p n}-1\right]^{-\gamma},\nonumber
\eea
with $\gamma=1/(s-1)$.
Therefore the fraction of nodes in the giant component can be approximated by
\bea
P_{\infty}(p)&\simeq &\exp\left[-\ln[\avg{k}(m-1)]a\int_0^{\infty}[x(n)]^2\right]\nonumber \\
&\simeq&\exp\left[-\mathcal {A} (\Delta p)^{-\sigma}\right],
\eea
where $\mathcal A$ is a constant and $\sigma={(s-3)}/{(s-1)}$.
Therefore the transition is continuous with a non-trivial singularity dictated by the exponent $\sigma$ which can in general be different from $\sigma=1$ and $\sigma=1/2$. This expression  reduces to the BKT singularity for $s=5$, i.e.,  $P_{\infty}(p)\propto \exp\left[-{\mathcal A}/\sqrt{\Delta p}\right]$ and  in the limit $s\to \infty$ reduces to the scaling $P_{\infty}(p)\propto \exp[-{\mathcal A}/\Delta p]$, but in general might be non-trivial. To our knowledge this anomalous critical scaling has not been reported previously  for any specific branching cell complex. 

Here we compare the RG predictions with extensive simulations for critical points of Eq.\ (\ref{FT}) of order $s=4,5,6$ (see Fig.\ \ref{fig.anomalous}). We have found evidence that the exponent $\sigma$ grows with the order $s$ of the critical point as predicted by the continuous RG approach. For $s=4$ we found a perfect agreement with the theoretical prediction $\sigma=1/3$. However for $s=5,6$ the exponent that we found numerically slightly differs from the predictions. This deviation from the theoretical predictions could be an effect of finite sizes or also due to the continuous approximation that we have used to predict the critical scaling.
 
\section{Conclusions}
In this work we have investigated the relation between network geometry and dynamics on branching simplicial and cell complexes.
Our main results are two-fold. On the one hand, we have shown that the discontinuous percolation transition is  observed not only in hyperbolic manifolds but also in branched non-amenable hierarchical networks. In this way we have generalized  previous results restricted to special cases of branching simplicial complexes.  Additionally, we have shown that, as the topology of the branched cell complex is evolving, the upper percolation transition can display non-trivial continuous critical behavior. On the other hand, we have shown that the considered non-amenable networks can have a number of intermediate phase transitions besides the upper and the lower one. At the lower percolation transition, the percolation probability becomes larger than zero but the giant component is not extensive. At the upper percolation transition the giant component becomes extensive. At the intermediate phase transitions the percolation probability and the fractal exponent have abrupt discontinuities but the fractal exponents remain smaller than one. Therefore below and above these intermediate phase transitions the giant component remains sub-extensive.
The latter result was derived by exploiting the mathematical similarities between the equations determining the percolation probability of simplicial and cell complexes with the equations determining the emergence of the MCGC in  correlated multiplex networks.
Despite the fact that the relation between the two percolation problems appears to  be only a formal one,  we expect that this result might be useful to further stimulate the research on the universal properties of explosive percolation problems. Namely, this work can be extended in several directions, including, for instance, the treatment of higher-dimensional simplicial and cell complexes and the case in which the $2$-dimensional cell complexes are formed by $m$-polygons with heterogeneous distribution for the number of faces $m$.  
\bibliography{biblio_cell2d}
\end{document}